# Customized Routing Optimization Based on Gradient Boost Regressor Model

Chen Zheng
Intel Corp.
Santa Clara, 95134

Clara Grzegorz Kasprowicz
Akm Semiconductor
San Jose, 95110

Carol Saunders
Akm Semiconductor
San Jose, 95110

**Abstract**
In this paper, we discussed limitation of current electronic-design-automoation (EDA) tool and proposed a machine learning framework to overcome the limitations and achieve better design quality. We explored how to efficiently extract relevant features and leverage gradient boost regressor (GBR) model to predict underestimated risky net (URN). Customized routing optimizations are applied to the URNs and results show clear timing improvement and trend to converge toward timing closure.

## 1. Introduction
Semiconductor technology node has been shrinking drastically, following Moore's Law [1]. For example, 7nm chip will be on market in 2017, and 5nm is in active research. The nanoscale transistor feature size as well as metallization minimum width and pitch brings in huge challenges to electronic-design-automation (EDA) tool in place and route space. Specifically in routing stage, extreme dense pin access requirement, high routing track congestions and rigorous design spacing rules made it impossibly difficult to close timing and design-rule-check (DRC). In such cases, generalized router function is not able to handle all types of situations well. Customized solution is needed to address hotspot, critical nets, etc. The effort spent to resolve these issues aggravate the burdens on designers and significantly hurt turn around time and design quality. In this paper, we propose a machine learning based framework that customize the routing characteristics on critical nets to improve routing quality. The approach overcomes the limitations of EDA tool and helps designers to achieve better quality of reference (QoR).

## 2. Related Works
Machine learning technique has been a hot research topic during recent years, it has wide application from image recognition [2] to natural language processing [3]. The machine learning application on integrated circuit design has not been vastly explored. For physical design optimization on router, most works focus on algorithm improvement [4], new feature integration [5] and design for manufacturability [6] or reliability [7]. There has been few works discussing applying machine learning technique to optimize routing on a design. In [8], A. B. Kahng discussed how to utilize machine learning technique to resolve congestion hotspot and reduce DRC count. In [9], J. Wuu discussed using machine learning for lithography hotspot detection. These works show some of the potential benefits by using machine learning technique to improve physical design quality. Thus, more exploration of machine learning in physical design is desired to help design in future technology.

In this paper, we focus on the router limitation mainly due to miscorrelation between global route and detailed route. In global route, router preliminarily estimate resistance-capacitance (RC) delay [10] of a net and roughly its cross-talk noise. Only timing-critical nets get assigned to higher layers or creating shield next to it to minimize RC delay or cross talk. However, due to the inaccuracy of delay estimation and unknown detailed routing results, global router often misinterpret a large number of actual critical nets. We refer to those nets as Underestimated Risky Nets (URN). Often, those nets will eventually end up routing on lower metal layers during detailed route and introduce significant delay impact on critical paths and result in negative slacks, which require substantial amount of manual effort for timing closure. It also causes lots of reliability problems as lower layer has smaller width [11].

The route engine algorithm can be improved to have better correlation between global route and detailed route; however, it is difficult to derive an analytical cost function to evaluate the actual net delay due to large number of varying parameters and unknown relationships between each parameter and evaluated result. Also, router algorithm aims to solve general cases, and globally applying settings may fix some issues but introduce other issues [12]. For example, setting layer effort to high may fix some of the layer misassignment issues but could possibly result in large count of additional DRCs. On the other hand, this situation falls perfectly into machine learning space, and is expected to generate promising results given powerful machine learn algorithms.

## 3. Feature Extraction

The most important factor to obtain a successful learning model is the input vector feature selection. Irrelevant features can inject lots of noise and train the model towards a random wrong direction. Due to technology node shrinking, many variations are also introduced into the routing space [13][14]. Via count is also a critical metric that can impact the net routing quality. Here, we determine that both timing and physical information is important in this case to predict routing degradation. Table 1 and 2 show possible relevant features we extracted on candidate nets and the final prunning of the features (gray out parameters are those filtered out according to feature importance).

Most machine learning package provide feature importance by training the model. Taking advantage of this, we are able to filter out low importance feature and only keep those relevant

| Table 1. Timing properties | | |
|---|---|---|
| path slack | net delay | setup time |
| arrival time | cross talk delay | skew |
| frequency | layer effort | uncertainty |

| Table 2. Physical properties | | |
|---|---|---|
| net length on each layer | via count on each layer | congestion on each layer |
| top layer | bottom layer | manhattan distance |
| No. fanout | driver strength | load capacitance |

features. In our experiments, we applied Gradient Boosting Regressor (GBR) from scikit-learn package in Python as the base learning model. The GBR model, can handle non-linear correlation between input vectors and output results as well as correlation between feature parameters [15]. Figure 1 shows the feature importances of a few parameters (some medium weight parameters are omitted due to limited space).

**Figure 1. Feature importance**

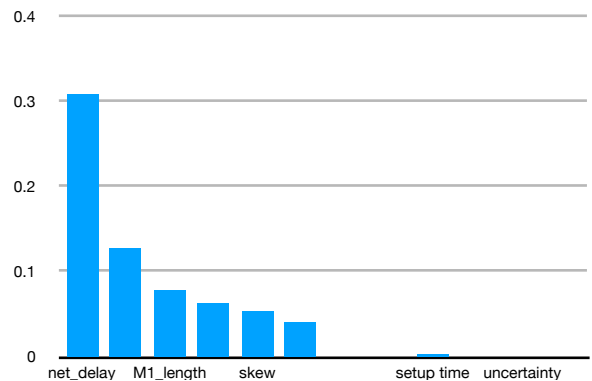

As expected, some of the features has no impact on the routing degradation, such as setup time, frequency, etc. However, there are also some counter-intuitive results. For example, M11 congestion has second highest importance. A typical mistake by data scientist is to filter out the counter-intuitive feature. This is absolutely a wrong operation and will significantly hurt the final result. According to machine learning theory, filtering out counter-intuitive feature based on subjective analysis is a serious error. Actually, the whole purpose of leverage machine learning technique is to help people find those hidden correlation behind the scene.

## 4. Training and Learning
In this section, we will discuss a few other factors that impact the learning results and how to apply the learning model to improve the design.

### 4.1 Training
#### 4.1.1 Sample size
Training sample size is a key factor to obtain successful training model. Over-training and under-training are two situations to avoid to generate better results. To achieve this, we follow the below criteria:

(1) Only nets on paths with slack less than 5ps are chosen as training samples;
(2) Only nets that preserved in both global route and detailed route are chosen as training samples.

Criteria (1) basically filters out non-critical nets, since router may route those nets differently and anyway those nets are safe and are of minimal impact on timing closure. Criteria (2) indicates that if a net gets buffered, there is no value of training these nets since router is already optimizing them. In our experiments, we randomly pick nets along 15K top critical paths among the top 30K critical paths as our training sample, and the nets along another random 5K paths among the remaining paths as validation set. Lastly, the remaining nets on the rest 10K paths are used as prediction set.

#### 4.1.2 Model parameters
The gradient boost tree model has a few key parameters to tune for best fitting the specific problem, including learning rate, n_estimators, subsample, min_samples_split, max_depth, alpha, max_leaf_nodes, etc. The parameter values are carefully tuned according to our problem size, unique features and subsequent constraints to produce best results for our experiments. The values are summarized in Table 3.

**Table 3. GBR Model Parameter values**

*learning_rate=0.2, n_estimators=200,*

*subsample=1.0, min_samples_split=4,*

*min_samples_leaf=1,*

*min_weight_fraction_leaf=0.0, max_depth=5,*

*alpha=0.9, max_leaf_nodes=None*

### 4.2 Learning
We train the model on an open source block or1200_if from OpenRISC 1200 [16]. The training model shows good prediction results on validation set as well as the prediction set. Results are shown in Table 4 and Table 5. Further, we apply the same model on a different block or1200_alu. The assumption is that the input vector features should be generally applicable independent of design, because using the input vector features (e.g. net_delay, layer_length, layer_congestion, etc.), one should be able to reproduce the net routing and with the given knowledge, router should behave consistently to route or optimize the nets. Table 6 shows the prediction results on the other block.

**Table 4. or1200_if validation set**

| #Total nets | 39748 |
|---|---|
| #Actual URNs | 2875 |
| #Positive prediction | 2372 |
| #Negative prediction | 95 |
| Positive rate | 0.825 |
| Negative rate | 0.033 |

**Table 5. or1200_if prediction set**

| #Total nets | 88497 |
|---|---|
| #Actual URNs | 8023 |
| #Positive prediction | 7157 |
| #Negative prediction | 209 |
| Positive rate | 0.892 |
| Negative rate | 0.026 |

**Table 6. or1200_alu 30K paths**

| #Total nets | 273871 |
|---|---|
| #Actual URNs | 28405 |
| #Positive prediction | 23747 |
| #Negative prediction | 1477 |
| Positive rate | 0.836 |
| Negative rate | 0.052 |

Once model is trained, designers can use this model and predict the router behavior on detailed route results based on global route information. Once prediction is done, designer can automatically apply different approaches to optimize the URNs. In our experiments, we choose two actions.: (1) layer assignment: user specified layer assignment can overwrite the tool default behavior and preserve the layer preference of URNs; (2) slack margin manipulation: if multiple URNs are detected along a common path, then path slack margin could be applied to give higher priority to the URNs. Routing on higher layers can give multiple benefits besides timing. For example, wires can tolerate more current [17] or achieve better reliability [18].

## 5. Experiments

We develop the proposed machine learning framework based on GBR package in Python. The design block is implemented using Synopsis ICC2 [19]. QoR results of baseline run and machine learning optimized run are summarized in Table 7 and 8.

**Table 7. or1200_if QoR**

| or1200_if | baseline | machine learning | improve rate |
|---|---|---|---|
| WNS (ns) | -0.174 | -0.155 | 0.112 |
| TNS (ns) | -415.866 | -264.907 | 0.363 |
| #violation path | 25391 | 17571 | 0.308 |

**Table 8. or1200_alu QoR**

| or1200_alu | baseline | machine learning | improve rate |
|---|---|---|---|
| WNS (ns) | -0.132 | -0.120 | 0.092 |
| TNS (ns) | -156.883 | -110.916 | 0.293 |
| #violation path | 8875 | 5973 | 0.327 |

Overall, worst negative slack (WNS) gets improved by an average of 10.2%, total negative slack (TNS) gets improved by an average of 32.8%, and number of violation path decreased by 31.8%. Figure 2 and Figure 3 show the net delay distribution shift of the URNs for or1200_if and or1200_alu for baseline and machine learning runs. The distribution curve clearly shows the net delay improvement by applying the proposed framework and a trend to converge towards timing closure.

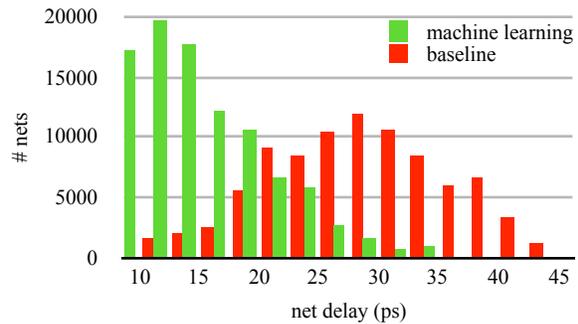

**Fig. 2. or1200_if net delay distribution shift**

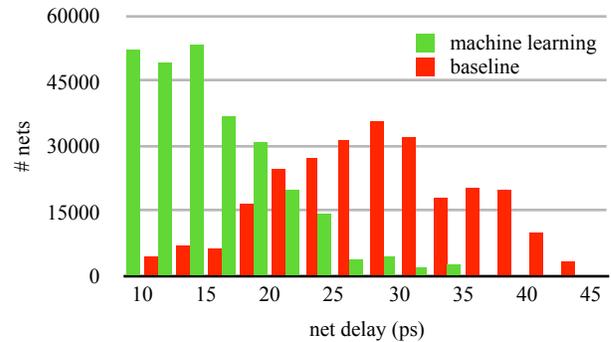

**Fig. 3. or1200_alu net delay distribution shift**

## 6. Conclusion

In this paper, we propose a new machine learning based approach to optimize routing during physical synthesis of an integrated circuit design. The proposed method can overcome router limits on miscorrelation between global router and detailed router. The results deliver significantly improved QoR with an average of 10+% WNS improvement and 30+% TNS improvement. It should be noted that such framework can be extended to a generalized platform for physical design, designers can extract any features they want and specify whatever patterns they want to classify or predict, and apply any optimization actions they want on the design. With help from machine learning technique, we expect to see a potential push to a higher performance from physical design space, as well as its applications on other physical design aspects, such as reliability, placement and clock synthesis [20-26].